\documentclass{iau} 
\usepackage{graphicx}
\usepackage{comment} 
\usepackage{graphics,graphicx,latexsym,amssymb,color}

\newcommand{\msun}{\ensuremath{\mathrm{M}_\odot}}

\newcommand\apjs{ApJS} 
\newcommand\apj{ApJ}
\newcommand\apjl{ApJ} 
 
\newcommand\araa{ARA\&A} 
\newcommand\mnras{MNRAS} 
\newcommand\nat{Nature} 
\newcommand\pasj{PASJ} 
\newcommand\aap{A\&A} 

\newcommand\prl{PRL}
\newcommand\pasa{PASA}

\title[SN Ia sub-classes \& progenitors] 
{Type Ia Supernova Sub-classes\\ and Progenitor Origin}

\author[Ashley J. Ruiter]   
{Ashley J. Ruiter$^1$}

\affiliation{$^1$School of Science, University of New South Wales Canberra \\
The Australian Defence Force Academy, 2600 ACT, Canberra, Australia \\ 
email: {\tt ashley.ruiter@adfa.edu.au} \\[\affilskip]
}

\pubyear{2020}
\volume{357}  
\jname{White Dwarfs as probes of fundamental physics and tracers of planetary, stellar \& galactic evolution}
\editors{NAMES OF EDITORS}
\begin{document}

\maketitle

\begin{abstract}
This paper presents a short review on the current state of SN Ia progenitor origin. 
Type Ia supernova explosions 
(meaning thermonuclear disruption of a white dwarf) are observed to be widely diverse in peak luminosity, lightcurve width and shape, spectral features, and host stellar population environment. In the last decade 
alone, theoretical simulations and observational data have come together to seriously challenge the long-standing paradigm that 
all SNe Ia arise from explosions of Chandrasekhar mass white dwarfs.
In this review I highlight some of the major developments (and changing views) of our understanding of the 
nature of SN Ia progenitor systems. 
I give a brief overview of binary star configurations and their plausible explosion mechanisms, 
and infer links between some of the various (observationally-categorized) SN Ia sub-classes and their progenitor origins from a theoretical standpoint.


\end{abstract}

\firstsection 
\section{Introduction}
 

It is accepted that Type Ia supernovae (SNe Ia) -- or thermonuclear supernovae -- are explosions of white dwarf (WD) stars in interacting binaries.  
When we talk about supernovae, SNe Ia are sometimes left out of the conversation since their progenitor population is poorly understood, unlike the case of core-collapse supernovae which are clearly understood to arise from the gravitational collapse of the cores in massive stars. For SNe Ia it is still unclear {\bf (1)} what type of companion star is donating matter to the WD and how this mass transfer proceeds (red giant or WD etc.; thermal timescale or dynamical timescale etc.), and {\bf (2)} what the WD mass (density) is when it explodes. This second point is most important for understanding what the explosion mechanism(s) is (are) in SNe Ia (e.g. \cite[Fisher et al. 2019]{Fisher2019a} and references therein). Understanding point (1) is necessary for knowing what ages of stellar populations give rise to SNe Ia and, characteristically, what observational properties the progenitor will exhibit. For progenitors with non-degenerate companions it might be possible to observe the binary system through pre-explosion observations (\cite[McCully et al. 2014]{McCully2014a}, see also \cite[Liu et al. 2015]{Liu2015a}), and it has been argued that in some cases, an excess flux would be produced soon after explosion due to SN ejecta-companion interaction (\cite[Bianco et al. 2011]{Bianco2011a}). On the other hand, with double degenerate progenitors it is challenging to detect such binaries without space-based gravitational wave observatories such as the future Laser Interferometer Space Antenna (LISA). With binary star evolution modelling it is possible to make predictions about which types of binary star systems are more likely to produce WDs that could explode as SNe Ia (birthrates), and what their characteristic ages (or `delay times') will be. Point (2), which needs to be investigated with sophisticated burning and explosion simulations, is crucial for informing which scenarios from part (1) are realistic, and for predicting observables such as ejecta velocities, abundance stratification, and nucleosynthetic yields (\cite[Thielemann et al. 1986]{Thielemann1986a}, \cite[Hillebrandt et al. 2013]{Hillebrandt2013a}). 

Despite a lack of a cohesive explanation for their progenitor origin, the role SNe Ia play in iron-group element nucleosynthesis, in driving a detailed understanding of interacting stars, and in determining cosmological parameters, is unequivocal.   
It has long been observed that the rate at which SN Ia lightcurves rose and declined was connected to the peak luminosity of the explosion (\cite[Arnett 1979]{Arnett1979a}, \cite[Phillips 1993]{Phillips1993a}), which itself was linked to the amount of $^{56}$Ni synthesized in the SN Ia (\cite[Pankey 1962]{Pankey1962a}). 
Though their explosion physics remains unclear, 
this information has nonetheless empowered SNe Ia to become our most useful objects for studying the expansion history (and future) of our Universe, and by extension, the nature of dark energy, which makes up most of the energy density of our Universe (Nobel prize in physics, 2011).
SNe Ia are also responsible for more than half of the iron in our own solar neighbourhood (\cite[Maoz \& Graur 2017]{Moaz2017a}), mostly from the nickel synthesized in the explosion through the radioactive decay chain $^{56}$Ni $\rightarrow$ $^{56}$Co $\rightarrow$ $^{56}$Fe. 
While very heavy elements (e.g. heavier than atomic number ${\sim} 40$) are primarily synthesized in low-mass stars during the late stages of evolution or compact object mergers, elements around the iron-peak (atomic numbers ${\sim} 22-30$, titanium to zinc) are created in supernovae; SNe Ia being dominant per event over core-collapse SNe. 

Since it is difficult to disentangle which progenitor scenario led to any hitherto observed SN Ia explosion,\footnote{This has been attempted extensively for the nearby SN 2011fe which exploded in M101 (\cite[R\"{o}pke et al. 2012]{Roepke2012a}).} and delay time distributions are intrinsically difficult to derive observationally especially for complex star formation histories, another useful constraint on SN Ia progenitors is nucleosynthetic yields. 
Manganese is an interesting element in this context because it is synthesized at high (white dwarf) densities, thus requiring at least some Chandrasekhar (or close to Chandrasekhar, which is ${\sim} 1.4$ \msun) mass WDs to have exploded as SNe Ia in the history of our Galaxy (\cite[Seitenzahl et al. 2013]{Seitenzahl2013a}). Otherwise, the solar abundance of manganese cannot easily be explained. 
Using different analysis methods, various authors have argued there is a requirement for a combination of Chandrasekhar mass and sub-Chandrasekhar mass progenitors, though the relative fractions do not usually agree (e.g. \cite[Scalzo et al. 2014]{Scalzo2014a}, \cite[Fl\"{o}rs et al. 2019]{Floers2019a}, see also \cite[Maguire et al. 2017]{Maguire2017a}).

\subsection{Diversity: What are the SN Ia progenitors?} 

Unlike core-collapse supernovae which are confined to exploding among young stellar populations, SNe Ia may explode as early as ${\sim} 40$ Myr up to (and beyond) a Hubble time after starburst. 
Figure 1 illustrates that while a large fraction of events thought to be of thermonuclear origin follow the Phillips relation (black curve; the empirical relationship between SN lightcurve width and peak luminosity that makes SNe Ia standardizable candles), a growing number (over $30$\%) of sources do not (see also recent review by \cite[Jha et al. 2019]{Jha2019a}). The term `normal' SN Ia is usually adopted to refer to a SN that tends to obey the Phillips relation and is therefore useful for cosmological studies; the SNe that do not follow this relationship are generally not used in such calculations (see e.g. \cite[D'Andrea et al. 2018]{D'Andrea2018a}). 

It has become very apparent that SNe Ia are a rather diverse group of objects, contrary to what was assumed over a decade ago. While it is becoming an accepted view that more than one progenitor scenario probably contributes (perhaps even to explain `normal' SNe Ia), the `progenitor problem' remains one of the longest-standing unsolved problems in stellar astrophysics (\cite[Maoz \& Mannucci 2012]{Maoz2012}). However, through a combination of numerical modelling (explosion simulations, radiative transfer models, binary star population synthesis) and observational data, it has been possible to make some progress toward solving this problem. 
It seems likely that the explosion originates in a WD of carbon-oxygen composition either through {\em delayed detonation} of a Chandrasekhar mass WD (deflagration-to-detonation transition, \cite[Khokhlov 1991]{Khokhlov1991a}), or a {\em prompt detonation} (e.g. via double-detonation) in a sub-Chandrasekhar mass WD (\cite[Woosley \& Weaver 1994]{Woosley1994a}). A few studies have shown that some thermonuclear events could plausibly arise from heavier, ONe WDs (\cite[Marquardt et al. 2015]{Marquardt2015a}, \cite[Jones et al. 2016]{jones2016a}), and there is a probable contribution from WDs that contain some non-negligible fraction of helium in the case of WD mergers (\cite[Perets et al. 2010]{Perets2010a}, \cite[Pakmor et al. 2013]{Pakmor2013a}, \cite[Crocker et al. 2017]{Crocker2017a}, \cite[Zenati et al. 2019]{Zenati2019a}). 
Table 1 shows a summary of the different labels attributed to sub-classes of the more peculiar SNe Ia in the literature and their main features from an observational standpoint (see Figure 1, which is adapted from \cite[Taubenberger 2017]{Taubenberger2017a}). The very luminous sub-class of `super-Chandrasekhar mass' explosions were proposed to originate from WDs that explode above the Chandrasekhar mass limit, though recent simulations have indicated that such systems, while indeed luminous, do not exhibit properties that are similar to this, or any other, sub-class (see \cite[Fink et al. 2018]{Fink2018a}). 
It is still not clear which explosions in Fig. 1 \& Table 1 arise from Chandrasekhar mass WDs (through delayed detonations) or sub-Chandrasekhar mass WDs (through prompt detonations).\footnote{Sometimes prompt detonations are referred to as `violent', e.g. violent WD mergers.} 
However, in a Chandrasekhar mass WD explosion, the WD may not always unbind because the (subsonic) deflagration fails to develop into a (supersonic) detonation. It is mostly well-accepted that this weak explosion can leave behind an `injured' WD as a remnant (\cite[Jordan et al. 2012]{Jordan2012a}, \cite[Kromer et al. 2013]{Kromer2013a}). Such an explosion of a CO or CONe WD is currently the leading explanation for the origin of SNe Ia of the ``Iax'' sub-class. Such left-over bound remnants are expected to exhibit unusual abundance signatures (\cite[Vennes et al. 2017]{Vennes2017a}). 
Numerous searches by different groups for donors of (presumably Chandrasekhar mass) supernovae have been mostly inconclusive so far (e.g. see \cite[Li et al. 2019]{Li2019a}), though possibly, donor stars that survived a SN Ia explosion from a sub-Chandrasekhar mass WD have already been found. Such donor stars in double degenerate systems that produce SNe Ia through a double-detonation -- if they survive -- will be flung out of orbit maintaining their pre-explosion velocity, which could be quite significant (on the order of $10^{3}$ km/s). \cite[Shen et al. 2018]{Shen2018a} found three such plausible donor stars in recent Gaia DR2 data, with one object exhibiting a likely path of origin that traces back to the Galactic supernova remnant G70.0-21.5. 

\begin{figure}[t]
\begin{center}
 \includegraphics[width=3.4in]{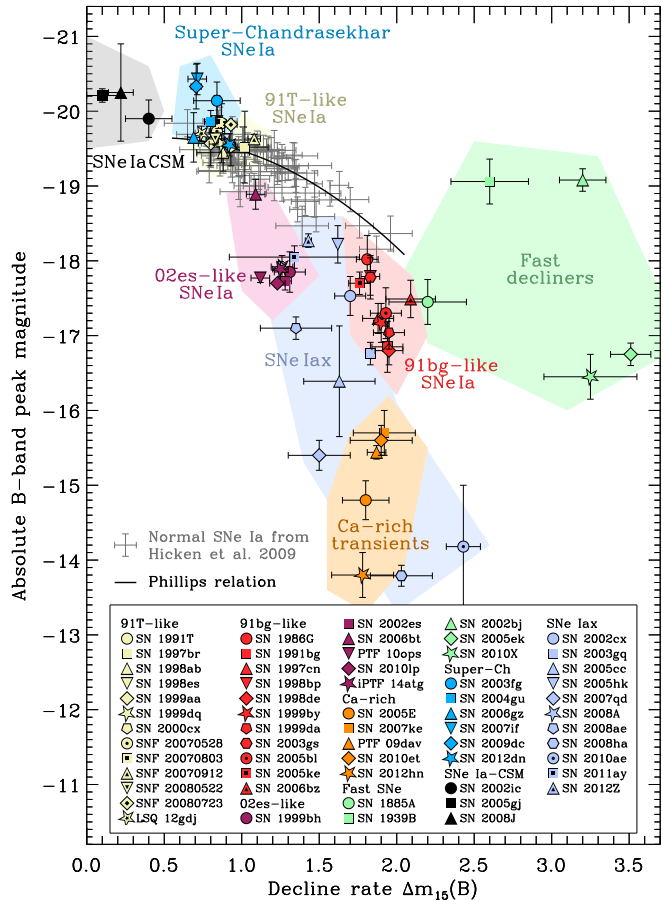} 
 \caption{Figure by S. Taubenberger showing various sub-classes of observed thermonuclear supernovae (see Taubenberger 2017).}
   \label{fig1}
\end{center}
\end{figure}

\begin{table}
  \begin{center}
  \caption{Synopsis of major SN Ia sub-classes (based on Figure 1 from Taubenberger 2017)}
  \label{tab1}
 {\scriptsize
  \begin{tabular}{|l|l|}\hline  
{\bf SN Ia sub-class } & {\bf observed characteristics} \\ \hline
  Ca-strong (previously `Ca-rich'): & high Ca to O nebular line ratios. Often far from potential galaxy host, see e.g. \cite[Shen et al. 2019]{Shen2019a}.\\ \hline
  1991bg-likes: & lower luminosity; faster than normal (e.g.\ narrow) light curves. \\ \hline
  1991T-likes: & higher luminosity; slower than normal (e.g.\ broad) light curves. \\ \hline
  super-Chandra: & high luminosity; low ejecta  velocities. Extra energy source unknown and possibly diverse. CSM? \\ \hline 
  SNe Ia-CSM: & high luminosity; thought to be due to interaction with circumstellar material (CSM). \\ \hline
  .Ia (``dot'' one A): & detonation of He-rich material on WD, preceded by weaker He-flashes; \cite[Bildsten et al. 2007]{Bildsten2007a}. \\ \hline 
  Iax: & diverse; low-luminosity, slow ejecta velocities, often young stellar pops; see e.g. \cite[Jha 2017]{Jha2017a}. \\ \hline
  2002es-likes: & sub-luminous with broad light curves (\cite[Ganeshalingam et al. 2012]{Ganeshalingam2012a}). \\ \hline
 fast decliners & Near-IR peaks similar to `normals'; possibly not thermonuclear origin (e.g. \cite[Drout et al. 2013]{Drout2013a}).\\ \hline
  \end{tabular}
  }
 \end{center}
\vspace{1mm}
\end{table}

Currently, the three leading binary star configurations, and possible matching explosion scenarios (see Section 4 for details), thought to lead to SNe Ia (including sub-classes) are:

\begin{enumerate}[(I)]
\item A Chandrasekhar mass white dwarf that is accreting matter in a single degenerate system (\cite[Whelan \& Iben 1973]{Whelan1973a}), where the white dwarf explodes successfully via delayed detonation (OR, the WD may not completely unbind in explosion if the initial deflagration does not transition into a detonation, a so-called `failed deflagration', or failed detonation). Multiple donor types are possible.   
\item A sub-Chandrasekhar mass WD that is accreting helium-rich material through RLOF and explodes below the Chandrasekhar mass limit via double-detonation (\cite[Livne \& Arnett 1995]{Livne1995a}). Donors are hydrogen-poor, helium burning stars, or white dwarfs that contain a non-negligible fraction of helium in their outer layers, either helium or HeCO `hybrid' WDs (\cite[Ruiter et al. 2014]{Ruiter2014a}). 
\item A double white dwarf merger, where mass transfer is unstable (dynamical timescale), and the explosion could be the result of either a Chandrasekhar mass white dwarf (\cite[Webbink 1984]{Webbink1984a}) through delayed detonation similar to (a), or a sub-Chandrasekhar mass WD likely via double-detonation similar to (b) (\cite[Pakmor et al. 2012]{Pakmor2012a}), but likley less helium is required. The latter is currently the favoured scenario in the double degenerate case. 
\end{enumerate}


\section{The path to simulating `Normal' SNe Ia} 

Until recently, theoretical predictions found CO+CO WD mergers were more likely to produce a neutron star via merger-induced collapse (MIC) rather than a thermonuclear explosion (see Section 4.1). \cite[Pakmor et al. 2010]{Pakmor2010a} was the first work demonstrating that the merger of two carbon-oxygen WDs, with roughly equal mass (${\sim} 0.9$ Msun each) could result robustly in a thermonuclear runaway in one WD that would rapidly unbind the star, burning to iron-group elements. The other WD would also be burned but would produce primarily lighter, intermediate-mass elements with little or no contribution to production of $^{56}$Ni, which also meant that the {\em peak brightness of the SN directly depended on the mass of the primary} (not the total binary mass). This study found however that these particular mergers did not produce enough $^{56}$Ni to look like normal SNe Ia, but their spectra and lightcurves matched fairly well to the sub-luminous 1991bg sub-class. It's important to note that in this pioneering study, the primary (exploding) WD underwent a single detonation in a sub-Chandrasekhar (0.9 solar mass) WD, not a double-detonation (see Section 4.2).

The realization in the last decade that SNe Ia could very plausibly arise from merging WDs in hydrodynamical simulations, and that the exploding WD did not have to approach the Chandrasekhar mass limit before detonating, has expanded the allowed parameter space for the SN Ia progenitor population. It has also called into question the viability of SNe Ia as standardizable cosmological candles.

\section{Evolutionary channels, SN Ia scenarios, and population synthesis} 

While detailed explosion model simulations and radiative transfer calculations of a single explosion configuration are critical for predicting synthetic observables, we also need to know how often said configuration would occur in Nature to explain the population that we see. 
It has long been assumed that all SNe Ia originated in a binary star system. It has however been hypothesized that a thermonuclear explosion 
may be possible in a single star within a specific mass range under the right conditions (a Type I 1/2 SN, \cite[Iben \& Renzini 1983]{Iben1983a}, see also \cite[Chiosi et al. 2015]{Chiosi2015a}). More recently, the idea that a non-negligible number of SNe Ia may arise in triple systems has gained some well-deserved attention (\cite[Di Stefano 2019]{DiStefano2019a}).  
Collisions of WDs (distinct from WD mergers) have also been proposed as a possible origin for SNe Ia (\cite[Kushnir et al. 2013]{Kushnir2013a}), but WD collisions are unlikely to account for the delay time distribution and rates of the SN Ia population (\cite[Toonen et al. 2018]{Toonen2018a}, \cite[Hallakoun \& Maoz 2019]{Hallakoun2019a}). 
The core-degenerate scenario (see early work by \cite[Sparks \& Stecher 1974]{Sparks1974a}, \cite[Livio \& Riess 2003]{Livio2003a}), in which a WD merges with the core of a giant-like star during a common envelope phase, has also been proposed as a possible scenario and has gained some recent attention (see review by \cite[Soker 2019]{Soker2019a}). So far such SNe Ia are mostly predicted to have short delay times (\cite[Wang et al. 2017]{Wang2017a}) and, contrary to the majority of SNe Ia, would likely show hydrogen in their spectra unless some mechanism prevented these systems from exploding soon after the merger begins (see also \cite[Ilkov \& Soker 2012]{Ilkov2012a}). While such mergers undoubtedly do occur in Nature and may produce interesting transients, to date, a lack of numerical simulations of this scenario precludes us from considering the core-degenerate scenario as a main formation channel for SNe Ia. 
Indeed, a number of formation channels could be contributing to the observed SN Ia population. The best way to quantify which types of stars produce SNe Ia, and in what relative numbers, is through rapid binary star evolution population synthesis (BPS). 

In a nutshell: BPS enables the rapid evolution of interacting stars through use of initial probability distribution functions for stellar orbital parameters ($a$, $e$, ZAMS masses and mass ratio $q$) and additionally incorporates treatment of the physical processes that should be taken into account in binary systems (tidal interactions, Roche lobe overflow [RLOF] mass transfer and accretion [including common envelope events], magnetic braking, gravitational radiation emission, supernova natal kicks, etc., see \cite[Toonen et al. 2014a]{Toonen2014a} for a BPS code-comparison study). What BPS lacks in detailed calculations of stellar structure and chemical composition (which can be investigated with detailed stellar evolution codes), it makes up for in rapid calculation of large populations of single and binary stars. 
While many star systems will not produce potentially explosive events, all `interesting' systems (e.g. potential SNe Ia) are flagged and can be studied in detail post-processing to uncover their evolutionary history, delay times, birthrates, and binary physical properties\footnote{Typically one BPS simulation involves the evolution from the ZAMS to a Hubble time with a single metallicity, but most BPS codes are capable of simulating stars with a wide range of metallicities.}.  
For BPS, initial distributions for orbital configurations were historically adopted using either a Salpeter or Kroupa IMF, the \cite[Abt 1983]{Abt1983a} initial separation, with a flat (meaning equal probability between 0 to 1) mass ratio distribution $q$ where $M_{\rm b}=qM_{\rm a}$ (with $M_{\rm b}<M_{\rm a}$), and a `thermal' distribution for initial eccentricities. More recently, especially when concerning stars that will end up as compact objects, it is reasonable to adopt initial distributions for orbital period, mass ratio and eccentricity from \cite[Sana et al. 2012]{Sana2012a} (see also \cite[Moe \& Di Stefano 2017]{Moe2017a}). 

The main scientific challenge in describing (numerically) interacting binary stars always stems back to how a star loses and/or gains mass (\cite[see e.g. Toonen et al. 2014b]{Toonen2014b}). This is a complex problem and has been tackled from a number of angles from different people over the years (e.g. \cite[Ritter 1988]{Ritter1988a}, \cite[van den Heuvel 1992]{vandenHeuvel1992a}, \cite[Hachisu et al. 1996]{Hachisu1996a}, \cite[Yaron et al. 2005]{Yaron2005a}, \cite[Nomoto et al. 2007]{Nomoto2007a}, \cite[Goliasch \& Nelson 2015]{Goliasch2015a}).  
Phases of unstable mass transfer involving at least one non-degenerate star with a distinct core-envelope structure are referred to as common envelope events. Such evolutionary phases are integral to SN Ia progenitors and are poorly-understood. Common envelope events are parameterized in BPS studies, with the possibility of adopting different formalisms and efficiency factors (\cite[Claeys et al. 2014]{Claeys2014a}). I will note that changing the efficiency ($\alpha$) or binding energy parameter ($\lambda$) does not appear to affect the overall rates of SN Ia progenitors too drastically; e.g. changing efficiency by an order of magnitude may only change birthrates by a factor of a few. This is because there is a shift, rather than an extreme suppression or enhancement, of progenitor parameter space. There is a noticeable effect on the types of progenitors that give rise to plausible SN Ia events. Detailed discussion of common envelopes was covered elsewhere in this symposium.   


\section{Type Ia supernova progenitor breakdown}

Below I give a brief summary of the two (delineated by explosion mechanism, not donor type) categories: Chandrasekhar mass and sub-Chandrasekhar mass explosions. 

\subsection{Chandrasekhar mass white dwarf SNe Ia}

 Traditionally, it was thought that SN Ia progenitors originated from either a `textbook' single degenerate, or (sometimes) a double degenerate, binary star system. It was believed that in either the case of a stably accreting WD (single degenerate) or in the case of a merger (double degenerate), the primary star (the more massive WD in the double degenerate case) would increase its mass via accretion, and would explode as it approached the Chandrasekhar mass limit (${\sim} 1.4$ M$_{\rm sun}$, e.g. once the central density achieved a high enough value such that carbon burning could take place; \cite[LeSaffre et al. 2006]{LeSaffre2006a}).  
In the single degenerate case when mass is transferred via RLOF, the donor can be a wide range of stellar types: a MS star, a sub-giant, a red giant, a hydrogen-stripped, helium-burning star (and possibly an AGB star in the case of efficient wind-accretion, \cite[Chiotellis et al. 2012]{Chiotellis2012a}). Chandrasekhar mass SNe Ia with hydrogen-poor, helium-burning donors will have very short delay times, and undergo different evolutionary phases, compared to Chandrasekhar mass SNe Ia with H-rich donors (see Section 5), even though the explosion mechanism in the WD is expected to be the same. 

It was found quite early on that if the entire 1.4 \msun WD exploded immediately as a detonation then the amount of $^{56}$Ni (and thus $^{56}$Fe) produced would be too high, being in tension with observations (\cite[Arnett et al. 1971]{Arnett1971a}). The `deflagration to detonation transition' was hypothesized to ameliorate this, in which the WD undergoes a period of expansion following sub-sonic carbon burning, leading up to the (super-sonic) explosion that unbinds the star. Since many SNe Ia observed before this century appeared to follow the Phillips relation, the notion that all explosions occurred from WDs with approximately the same (Chandrasekhar) mass, thus producing the same mount of $^{56}$Ni that would power the lightcurve, seemed to be a natural consequence. Note that in this explosion scenario carbon is ignited once the central density of the WD attains a critical value, but the star only explodes ${\sim}$centuries after this `simmering' phase of carbon burning.  

Early theoretical studies of merging WDs indicated that rapid accretion on the primary would not result in a SN Ia since off-centre carbon burning was more likely to occur (\cite[Yoon et al. 2007]{Yoon2007a}), thereby enabling the carbon-oxygen WD to escape an explosive fate and rather evolve into an oxygen-neon-rich WD. An ONe WD is more likely to collapse to form a neutron star via accretion induced collapse (\cite[Miyaji 1980]{Miyaji1980a}, \cite[Saio \& Nomoto 1985]{Saio1985a}) when it approaches the Chandrasekhar mass (but see \cite[Jones et al. 2016]{Jones2016a}). 

As mentioned previously, searches for stellar companions have focused on finding possible donors in the Chandrasekhar mass scenario, since until very recently, it was thought that double WD progenitors would leave behind no companion star. While searching for companions from single degenerate scenario explosions has not led to any definitive discovery of a progenitor configuration, observations of ionized gas in the vicinity of SN Ia remnants is another way of constraining the nature of the donor star: The pre-supernova luminosity (and thus stellar type) of the donor can be inferred from the level of hydrogen and/or helium ionization experienced in nebular region of the remnant (\cite[Woods et al. 2018]{Woods2018a}, \cite[Kuuttila et al. 2019]{Kuuttila2019a}). Thus far, the main conclusion is that in most remnants studied, the progenitor system could not have contained a hot, luminous donor which we could expect in the presence of a nuclear-burning WD undergoing accretion from a non-degenerate companion. In other words: a `textbook' Chandrasekhar mass progenitor is disfavoured for these remnants.


\subsection{Sub-Chandrasekhar mass white dwarf SNe Ia}

Studies in the 1990s investigated the possibility of sub-Chandrasekhar mass SNe Ia, in which a carbon-oxygen WD could explode at a mass well below 1.4 \msun \, under the right conditions: the WD needed to be accreting matter that was helium-rich (\cite[Woosley \& Weaver 1994]{Woosley1994a}, \cite[Livne \& Arnett 1995]{Livne1995a}).\footnote{The sub-Chandrasekhar mass scenario was also explored in the context of symbiotic binary systems (\cite[Yungelson et al. 1995]{Yungelson1995a}). } At fairly low (${\sim}$few $\times 10^{-8}$ \msun/yr) accretion rates of helium, a WD can build up a thick helium shell; the low accretion rate suppresses the effect of helium nova flashes (e.g. \cite[Kato et al. 2008]{Kato2008a}). 
Eventually the base of the layer can become degenerate and ignite.
This helium-shell detonation can be successful in driving a second detonation close to the WD core (\cite[Shen \& Moore 2014]{Shen2014a}), thus the term `double-detonation'. However, early work showed that the thick ${\sim} 0.2 $ \msun \, layer of helium shell needed to initiate the detonation was in severe tension with observations of SN ejecta; the models over-produced iron-peak elements at high velocities. It was shown only over the last decade or so (\cite[Fink et al. 2010]{Fink2010a}, \cite[Woosley \& Kasen 2011]{Woosley2011a}) that the helium shell did not need to be so massive; only on the order of 5\% of a solar mass, for a shell detonation to proceed (the shell mass at detonation is dependent on the WD mass, \cite[Piersanti et al. 2015]{Piersanti2015a}). With the decrease in the required shell mass, theoretical lightcurves and spectra showed an improved match to observations (\cite[Kromer et al. 2010]{Kromer2010a}) -- on par with what was being found for the WD merger and Chandrasekhar mass scenarios. Thus, the double-detonation model was rejuvinated. 
Interestingly, recent work has shown that explosions from sub-Chandrasekhar mass WDs accreting helium in this `classical' double-detonation scenario (e.g. helium accretion rates on the order of ${\sim}$few$ \times 10^{-8}$ Msun/yr) do not exhibit properties of normal SNe Ia but may produce rapid transients, possibly events of the Ca-strong sub-class (see \cite[Neunteufel et al. 2017]{Neunteufel2017a}, which includes effects of magnetic fields and WD rotation).

Historically, WD mergers were not expected to produce SNe Ia very readily. However, since the 1980s, the idea that WD mergers could make thermonuclear supernovae started to gain ground (\cite[Webbink 1984]{Webbink1984a}, \cite[Iben \& Tutukov 1984]{Iben1984a}). These SN Ia scenarios of WD mergers proposed decades ago generally presumed that during the merger, the secondary WD would become disrupted, form an accretion torus around the primary, and the primary would accrete matter and eventually (after thousands of years) explode as it approached the Chandrasekhar mass limit. WD mergers that did not lead to a primary WD approaching this mass limit could form R Coronae Borealis stars, sdO stars, and other objects (\cite[Ruiter HDEF talk 2018]{Ruiter HDEF talk 2018}). While as discussed in 4.1, this `double degenerate merger to Chandrasekhar mass' scenario may indeed be plausible, WDs that explode by unstable mass transfer are now expected to do so early in the merging process, while still at sub-Chandrasekhar mass. After the \cite[Pakmor et al. 2010]{pakmor2010a} study that demonstrated this was numerically viable in three-dimensional WD merger simulations, other groups began exploring the parameter space more extensively (see \cite[Schwab et al. 2012]{Schwab2012a}, \cite[Dan et al. 2014]{Dan2014a}, \cite[Raskin et al. 2014]{Raskin2014a}, \cite[Zhu et al. 2015]{Zhu2015a}, \cite[Sato et al. 2016]{Sato2016a}). \cite[Guillochon et al. 2010]{Guillochon2010a} demonstrated that WD mergers involving a helium-rich secondary could successfully form a detonation in the primary WD's accreted helium-rich envelope, which could possibly lead to a detonation in the primary CO WD. The idea that the presence of helium in a WD merger could facilitate the detonation process in a sub-Chandrasekhar mass WD was another important leap forward, enabling sub-Chandrasekhar mass WDs to explode more easily through double-detonations. It was proposed by \cite[Pakmor et al. (2013)]{Pakmor2013a} that CO+CO and CO+He-rich WD mergers might form a continuum of progenitors, accounting for both normal and sub-luminous events. It is expected that so-called `pure' CO WDs would also contain some helium on their surfaces left over from previous stellar evolution, and this small amount of helium could possibly be sufficient to enable the first (surface) detonation in the double-detonation process during unstable mass transfer. It was demonstrated with AREPO (\cite[Springel et al. 2010]{Springel2010}) simulations consisting of CO WDs in Pakmor et al. 2013) that a helium detonation in a thin helium shell on the surface of the primary WD could be ignited during a merger. This opened the door to not only facilitating the thermonuclear explosion mechanism for sub-Chandrasekhar WDs, but also meant that it was now easier for WD mergers to produce `normal' SNe Ia as well as other sub-classes, with the exploding (primary) WD mass being the main determinant for absolute peak luminosity through production of $^{56}$Ni.
   In a study that combined binary population synthesis, 1D hydrodynamic explosion models and synthetic observables through radiative transfer calculations, it was shown that the peak luminosity distribution of local (within 80 Mpc) SNe Ia could indeed be reproduced by sub-Chandrasekhar mass CO WDs undergoing a merger. It was revealed that nearly half of the progenitors end up going through an extra phase of RLOF mass transfer long before the formation of the secondary WD (\cite[Ruiter et al. 2013]{Ruiter2013a}, fig. 2). It is this extra mass transfer stage that leads to a shift of the average primary WD mass toward higher masses in WD merger progenitors. This primary WD mass shift is responsible for bringing the predicted peak brightness distribution into agreement with observations (\cite[Ruiter et al. 2013]{Ruiter2013a}, figs. 5 and 8).

\section{Rates and delay times}

While for decades astronomers have known about real binary star systems that fit the profile for the textbook single degenerate scenario of SNe Ia (e.g. recurrent novae; see \cite[Hillman et al. 2016]{Hillman2016a} for a modern review of such systems), many population synthesis studies have found that it is difficult to substantially increase the mass of a CO WD via hydrogen accretion, though some optimistic prescriptions for mass accretion efficiency and retention can bring up the potential rates in this scenario (see e.g. \cite[Nelemans et al. 2013]{Nelemans2013a}). The theoretically-predicted rate of SNe Ia from the `textbook' single degenerate Chandrasekhar mass channel still appears to fall rather short compared to the theoretically-predicted rates of double degenerate mergers. While delay time distributions of WD mergers from BPS models are canonically found to have a mostly smooth (beyond a few 100 Myr) shape, the single degenerate Chandrasekhar mass and sub-Chandrasekhar mass delay time distributions are each bimodal: those with hydrogen-poor, helium-burning donors occur at short delay times (see \cite[Hillebrandt et al. 2013]{Hillebrandt2013a}). 

An observational study of the delay time distribution of SNe Ia\footnote{The time at which a supernova occurs relative to its ZAMS formation time; applicable to coeval star systems.} in old stellar populations by \cite[Totani et al. (2008)]{Totani2008a} demonstrated that this distribution followed a power law shape with ${\sim} t^{-1}$. A distribution of delay times that follows such a power-law is expected if the initial distribution of binary stars even roughly follows an initial distribution where the separation is flat in log (see \cite[Ruiter et al. 2009]{Ruiter2009a}). This type of delay time distribution shape is not especially expected for single degenerate progenitors where the delay time is primarily governed by the evolutionary timescale of the donor rather than a decrease in orbital period due to gravitational wave radiation. By its nature however, the Totani et al. study could not probe delay times shorter than ${\sim 0.2}$ Gyr. Measuring the delay time distribution at short delay times is indeed more challenging, since we have to either be successful in studying high-redshift SNe (\cite[Friedmann \& Maoz 2018]{Friedmann2018a}), or re-construct the SN delay time from supernova remnants of Type Ia in local galaxies with active star-formation (\cite[Maoz \& Badenes 2010]{Maoz2010a}). It is still unclear why the rate of SNe Ia in galaxy clusters is measured to be higher than the rate derived from field galaxies (see \cite[Maoz \& Graur 2017]{Maoz2017a}, fig. 2), though metallicity may play some role (see Summary).

\begin{figure}[t]
\begin{center}
 \includegraphics[width=3.1in]{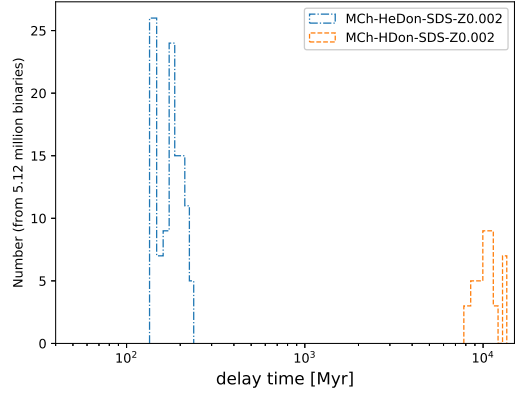} 
 \caption{Theoretical delay time distribution for single degenerate, Chandrasekhar WD mass SNe Ia computed with {\tt StarTrack} assuming a metallicity of Z=0.002 (${\sim} 10$\% solar). Accretion is through RLOF only (no wind accretion assumed). Explosions with H-stripped, He-burning star donors have short delay times while those with H-burning star donors more typically have longer delay times.}
   \label{fig2}
\end{center}
\end{figure}

\begin{figure}[t]
\begin{center}
 \includegraphics[width=3.1in]{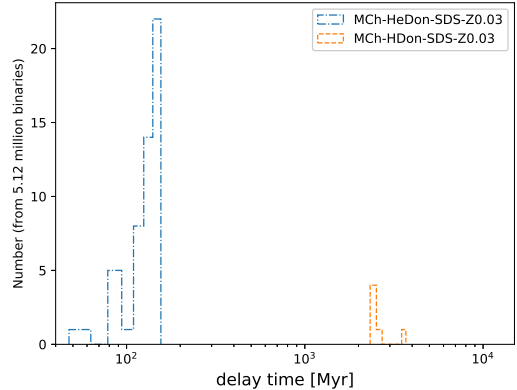} 
 \caption{Theoretical delay time distribution for single degenerate, Chandrasekhar WD mass SNe Ia computed with {\tt StarTrack} assuming a metallicity of Z=0.03 (${\sim} 50$\% above solar). Accretion is through RLOF only (no wind accretion assumed). Explosions with H-stripped, He-burning star donors have short delay times while those with H-burning star donors more typically have longer delay times.}
   \label{fig3}
\end{center}
\end{figure}

\subsection{Metallicity} 
Preliminary studies of the impact of ZAMS metallicity on SN Ia progenitors have shown that initial metallicity can play an important role regarding which types of evolutionary channels for SNe Ia may be favoured at lower vs. higher metallicity.   
It is probably wise to keep in mind that using lower redshift (higher metallicity) SNe Ia to standardize SNe Ia at large redshift (lower metallicity) in order to perform precise calculations of cosmological quantities may then lead to difficult-to-trace uncertainties in the derivation of cosmological parameters. 

Figures 2 and 3 show the delay time distribution from {\tt StarTrack} (\cite[Belczynski et al. 2008]{Belczynski2008a}, \cite[Belczynski et al. 2002]{Belczynski2002a}) for Chandrasekhar mass WD explosions originating from binaries in RLOF where the donor was either H-rich (typically a red giant or sometimes a star in the Hertzsprung gap), or a H-stripped, He-burning star (typically a helium sub-giant). The systems involving H-stripped, He-burning stars arise from more massive progenitors that leave the main sequence rather quickly, and tend to always have delay times ${\lesssim} 300$ Myr. The formation of such a progenitor normally involves two common envelope events and one phase of stable mass transfer. In the case of the `textbook' H-rich donor Chandrasekhar mass channel, the formation of such a progenitor usually involves only one common envelope event and one stable phase of mass transfer. It should be noted that a trend was found in the {\tt StarTrack} data where at very low metallicities (e.g. Z=0.0001), the rate of SNe Ia from H-rich donors can become suppressed because it is more favourable to produce ONe WDs rather than CO WDs due to the decreased stellar wind mass-loss rates. 



Changing the initial metallicity will affect the stellar core mass and thus will influence the birth mass of the WD. 
The H-poor, He-burning donor systems with delay times above 200 Myr do not occur in the higher Z model because the same binary would undergo a different evolutionary pathway on account of the smaller core (and hence WD) mass. The change in mass (and mass ratio) not only affects the chance of the WD reaching the Chandrasekhar mass limit, but is complicated by the fact that a given binary configuration may or may not reach contact and undergo stable RLOF, and if contact is achieved, that mass transfer may or may not be stable (see \cite[C\^{o}t\'{e} et al. 2018]{Cote2018a} for an analagous example involving nova progenitors). 
For example, a typical system evolved with Z=0.002 is not likely to produce a SN Ia if it is evolved with Z=0.03. Even if the higher-Z system could result in a RLOF binary within a Hubble time, the lower core mass (due to increased wind mass loss rates) will inhibit formation of a Chandrasekhar mass WD. 
On the other hand, for the same binary evolved from very low-Z (Z=0.0001), the primary star is more likely to evolve into an ONe WD than a CO WD, possibly resulting in an increased number of accretion-induced collapse neutron stars relative to Chandrasekhar mass SNe Ia (\cite[Ruiter et al. 2019]{Ruiter2019a}). 
 
The systems with H-stripped, He-burning donors -- if they fail to produce a detonation after ignition -- are good candidates for explaining some members of the SN Iax population, such as SN 2008ha (\cite[Foley et al. 2014]{Foley2014a}, \cite[Kromer et al. 2015]{Kromer2015a}) or possibly SN 2012Z (\cite[McCully et al. 2014]{McCully2014a}).

\section{Summary}

In what ways can we constrain the theoretically-predicted scenarios of SNe Ia, and match these to various members of the SN zoo? In terms of progenitor clues, the merger scenario progenitors discussed in Ruiter et al. 2013 predict a phase of binary evolution where the first-formed WD undergoes a short period (${\lesssim} 1$ Myr) of accretion of helium-rich material long before the double WD is formed. This phase of He-rich accretion would result in a helium nova system. If such progenitors are contributing significantly to the SN Ia population then we would expect on the order of a few to 10 of such He novae in the Galaxy today (V445 Pup is such a system). With the launch of LISA projected to be in 2034, we can look forward to unprecedented data from double WD binaries that will give us further insight into the formation and evolution of these systems. Currently, WD mergers appear to tick many of the boxes in explaining SNe Ia: delay time distribution, lack of hydrogen, peak brightness distribution, explosion mechanism, and the predicted rates are on par with the empirically-derived `Galactic' SN Ia rate (though usually fall a factor of a few short in explaining rates in galaxy clusters).
A subset of He+CO WD mergers are predicted to have rather long delay times, and if these binaries undergo helium detonations, they could explain the Galactic antimatter signal (\cite[Timmes et al. 1996]{Timmes1996a}, \cite[Crocker et al. 2017]{Crocker2017a}) and account for the 1991bg SN sub-class (\cite[Panther et al. 2019]{Panther2019a}).   
For SNe Ia of Chandrasekhar mass, the predicted rates remain lower than those of double degenerate mergers, though the `failed deflagration' (or rather failed detonation) model involving a H-stripped, He-burning star donor seems very promising in explaining the faint thermonuclear SNe of the Iax sub-class. Regarding accretion of helium-rich material -- the `classical' double-detonation scenario in which a sub-Chandrasekhar mass WD accretes stably from a He-rich WD or a H-stripped, He-burning star: it is still not clear what range of He shell masses is acceptable if these systems have a hand in contributing to normal SNe Ia. SN 2018byg was a peculiar, rare transient thought to originate from a double-detonation that is predicted to have contained a rather thick helium shell (\cite[De et al. 2019]{De2019a}). 
A recent study by \cite[Seitenzahl et al. (2019)]{Seitenzahl2019a} used MUSE integral field spectroscopy data of SN Ia remnants to interpret the nature of these explosions. Using optical coronal lines of [Fe XIV] to constrain their shock models, they found that in one case the progenitor was best-explained by a Chandrasekhar mass WD, while in another case it could be best-explained by a WD of sub-Chandrasekhar mass. This exciting result has given the supernova community a new method for revealing the origin of SNe Ia. 

A general trend found with StarTrack BPS data is that SN Ia efficiency decreases with metallicity (more SNe at lower Z). This trend is not yet fully understood in the context of binary star evolution, but it does go the right way considering cosmological SN Ia rates are observed to be higher than local field rates. Separately, recent observations also indicate that binary fraction (of solar-type stars) goes up as Z goes down (\cite[Moe et al. 2019]{Moe2019a}). To first order this implies that there is more opportunity for stellar interactions at lower Z which may account for an increase in the SN Ia rate at lower Z. In any case, The BPS models investigated here employed the same binary star fraction at all metallicities, so there are likely to be additional metallicity effects that play important roles regarding birthrates (e.g. wind mass loss rates affecting mass transfer; stellar core and hence WD mass). Indeed metallicity seems to play some role in SN Ia evolution and/or progenitor formation, but pinpointing all of the reasons how is still a work in progress. 

In the coming years, surveys such as ZTF and LSST will find more supernovae. Detailed observations of a handful of nearby SNe will be more useful in constraining the progenitor nature than having a large (possibly incomplete) sample of faint events. But explaining the progenitor of any one explosion will be difficult unless a pre-explosion progenitor is confirmed, the explosion occurs nearby, or, in the case of SN remnants, is within a certain age range. Despite those unlikelihoods, it is becoming possible to set limits on what types of SNe exploded among ancient stellar populations. With the recent advances in theoretical yield predictions from different SN Ia explosion models, it is within our reach to delineate between different explosion scenario histories in small galaxies with relatively simple star formation history (\cite[Venn et al. 2016]{Venn2016a}, \cite[Cescutti \& Kobayashi 2017]{Cescutti2017a}, \cite[Kirby et al. 2019]{Kirby2019a}).

\section{Acknowledgments}

AJR acknowledges financial support from the Australian Research Council under grant FT170100243, and thanks the SOC of the IAUS 357 for the invitation. 
AJR also thanks Stefan Taubenberger for the updated plot (Figure 1), and thanks Ivo Seitenzahl for some useful references. 
This research was undertaken with the assistance of resources and services from the National Computational Infrastructure (NCI), which is supported by the Australian Government, through the National Computational Merit Allocation Scheme and the UNSW HPC Resource Allocation Scheme.
AJR is thankful for the New South Wales Rural Fire Service, who are mostly volunteers, for their hard and important work. They were constantly in my thoughts while I was finishing this paper in the early days of 2020, while Australia entered its worst-ever bushfire season.

\end{document}